\magnification 1200 \baselineskip 20pt 
\vbox{\vskip 2cm} \rightline{CINVESTAV-94/06} \rightline{May 1994}

\vskip 2truecm

\centerline{\bf On Self-dual Gravity Structures on Ground Ring }
\centerline{\bf Manifolds in Two-Dimensional String Theory }

\vskip .5truecm

\centerline{ H. Garc\'{\i}a-Compe\'an\footnote*{Supported by a CONACyT
Graduate Fellowship.}, J.F. Pleba\'nski\footnote{\dag}{On leave of absence from
the University of Warsaw, Warsaw, Poland.} and M.
Przanowski\footnote{\ddag}{Permanent address: Institute of Physics, Technical
University of L\'od\'z, W\'olcza\'nska 219, 93-005 L\'od\'z, Poland. Supported
by CINVESTAV and CONACyT.}}

\centerline{\it Departamento de F\'{\i}sica, Centro de Investigaci\'on y
de Estudios Avanzados del I.P.N.}

\centerline{\it Apdo. Postal 14-740, 07000, M\'exico D.F., M\'exico}
\vskip 1truecm
\centerline{\bf Abstract}

We generalize the geometric structures generated by  Witten's ground ring.
 It is shown that these generalized structures involve in a
natural way some geometric constructions from Self-dual gravity [1,12].
The
formal twistor construction on full quantum ground ring manifold is also
given.


\vfill
\break

\centerline {\bf 1. Introduction}

In recent years, the search for a physical principle to be represented by
some mathematical structure of String Theory has become essential.  In
this direction, a most interesting proposal states that this theory
should be background independent. This idea has been discussed in a
number of papers by Witten and Zwiebach [2,3]. A Lagrangian
which is invariant under changes of backgrounds (just as the Yang-
Mills Lagrangian is invariant under gauge transformations) is proposed in a
2-dimensional string toy model. The present work deals with an element of
"the space of all 2-dimensional field theories" [2].

Quantum
gravity in two dimensions is a system of integrable models in which
some quantities are computed which are very difficult to find in four
dimensions. Liouville theory of 2d gravity via discrete matrix models
is an example of these
with central charge $c\leq 1$. These models can be characterized for the case
$c=1$ with usual decoupling between "matter" and ghost as the background

$$ 2D\ Quantum\ Gravity \otimes CFT\otimes ghost. $$
It is very important to study the physical states for these
backgrounds (including their symmetries) in order to understand
the structure of the vacuum in 2-dimensional string theory [4,5]. It has
been discovered an
infinite number of discrete states with spin zero and ghost number
equal to one.

Witten proved that states with spin zero and ghost number zero
have in a natural way the mathematical structure of a commutative
ring with generators (in the uncompactified SU(2) point) ${\cal O}_{u,n}$ with
$u=s-1\in \{0,{1\over 2},1,{3\over 2},...\}$ and $n\in \{u,u-1,...,-u\}$
[4] (for details and a short summary of the ground ring structure see
Appendix). Here we have the generators

$$ {\cal O}_{0,0} \equiv 1 \eqno(1.1)$$
for $s=1,\ u=0,\ n=0$ and

$$ x \equiv {\cal O}_{{1\over 2}, {1\over 2}} = (cb + {i\over 2}(\partial X - i
\partial \phi)) exp [i(X + i \phi)/2] $$

$$ y \equiv {\cal O}_{{1\over 2}, -{1\over 2}} = (cb - {i\over 2}(\partial X +
i \partial \phi)) exp - [i(X - i \phi)/2] \eqno(1.2)$$ for $s={3\over 2},
u={1\over 2}, n={1\over 2},-{1\over 2}$. Under the product given by the
operator product expansion, the states generate a structure called {\it Chiral
Ground Ring}. The states encode all relevant information about the symmetries
of the theory, and furthermore they give rise to recursion relations
among the
tachyon bulk scattering amplitudes perturbing the ground ring [6,7]. The
symmetries involved here are of two types:

\item{a)-.} The group $SDiff({\cal A})$ of diffeomorphisms preserving the area
of a flat two-dimensional ring manifold. The coordinate functions of ${\cal
A}$ are precisely the chiral ground ring generators, ${\cal A} = \{$x,y$\}$.
In other words, if $\omega$ is a volume form on ${\cal A}$ {\it i.e.} $\omega
\in \Omega ({\cal A}),\ \ \omega = dx \wedge dy$ we have that if $\phi \in
SDiff({\cal A})$, then $\phi^*\omega = \omega$.

\item{b)-.} In the case of non-chiral ground ring (or the full quantum
ground ring) we have a combination of a left and right movers represented by
the ground rings, $C({\cal A}_L)$ and $C({\cal A}_R)$ respectively, where
${\cal A}_L = \{x,y\}$ and ${\cal A}_R = \{\bar x, \bar y \}$. This
combination is needed to form the spin $(0,0)$ quantum field operators ${\cal
V}_{u,n,n'} = {\cal O}_{u,n} \cdot \bar {\cal O}_{u,n'}$ and can be considered
to be a tensorial product of rings $C({\cal W}) = C({\cal A}_L) \otimes
C({\cal A}_R)$ with generators  $ x,y,\bar x, \bar y$. Again these generators
are the coordinate functions on certain flat four-dimensional manifold ${\cal
W}$. In a like manner there is a symmetry group ${\cal H}$
acting on ${\cal W}$; this
action consists in identifying the generators of the form $x,y,\bar x,\bar y
\rightarrow tx,ty,t^{-1}\bar x,t^{-1}\bar y$. The group  ${\cal H}$ is
generated by the vector field $ S = x{\partial \over \partial x} +
y{\partial \over \partial y} - \bar x {\partial \over \partial \bar x} - \bar
y{\partial \over \partial \bar y}$. The quotient space ${\cal W}/{\cal H} =
{\cal Q}$ has the topology of a 3-dimensional cone. The symmetry group that
generates the states in 2-dimensional string theory is just the group of
symplectic diffeomorphisms preserving the volume of this cone. The coordinates
of the cone have a natural interpretation in the context of $c=1$ matrix
models. [ Note that the group of diffeomorphisms preserving a 3-dimensional
volume has been considered by Takasaki [8] in the context of infinite
hierarchies, within  Pleba\'nski's approach to 4d self-dual gravity [9]]. Very
recently Lian and Zuckerman [10] have shown that for every string background
the BRST cohomology has an algebraic structure known as {\it Gerstenhaber
algebra} and they suggest that the homotopy Lie algebras of the Gerstenhaber
algebra and the algebra constructed by Witten and Zwiebach [5] are closely
related.

There exists a nice geometric differential approach to this set of things
on the real flat 4-dimensional ring manifold ${\cal W}$ as well as on the 3-
dimensional algebraic variety ${\cal Q}$. In this paper we intend to make a
natural generalization of the differential geometric methods in 2d string
theory by considering, instead of ${\cal W}$, a Ricci half-flat ({\it i.e.}
$R_{\dot A \dot B} =0$) K\"ahler 4-ring manifold ${\cal M}$, with a chart
given
by  the local complex coordinate functions  $\{x^{\mu}\} = \{x,y,\bar x,\bar
y\}$. Then an open set of ${\cal M}$ will thus look like the
complexification of ${\cal
W}$, {\it i.e.} ${\cal W}^{\cal C}$. (Notice that we are using the same letter
to designe the general manifold and the ${\cal C}^4$-plane).  ${\cal W}^{\cal
C}$ is the product of two complex 2-ring manifolds: ${\cal W}^{\cal C} =
{\cal
A}^{\cal C}_L \times {\cal A}^{\cal C}_R$. It is remarkable that ${\cal
A}^{\cal C}_L$ and ${\cal A}^{\cal C}_R$ are symplectic manifolds with
symplectic 2-forms given
 by $\omega$ and $\tilde \omega $ respectively. This fact causes that,
automatically, many of the constructions  given in [1,9,11,12] lead to
the self-dual gravity structures in two dimensional string theory.

In Section 2 we briefly review some basic results of the Witten-Zwiebach
theory in terms that can be useful for the search of the self-dual
gravity in string theory.

Section 3 is devoted to study self-dual gravity [9] in the
context of 2d string theory. We claim that the Witten-Zwiebach theory finds a
natural generalization within the ${\cal H}$ and ${\cal H-H}$ spaces theory
[13,14]. It is also demonstrated that our construction appears to be
exactly that of Witten-Zwiebach locally.

In Section 4, we  prove  the existence of self-dual gravity
structures on the full quantum
ground ring manifold in a different way as that given by Ghoshal {\it et
al}. This new method gives a deeper understanding of their results [15].
Using the construction of Ref.[1] we
re-derive the first heavenly equation from the natural symplectic
structure on the
chiral ground ring manifold. Then, we make some comments about the
curved twistor construction on the ground ring manifold. We first
identify the chiral
ground ring manifold with the twistor surface, and then show
for the ground ring manifold that any local information (relevant to 2d
string theory) can be completely represented on some twistor space.

    Finally, in Section 5  some further implications of this work are
considered.

\vskip 2truecm

\centerline{\bf 2. Generalities and Ground Ring Manifolds}
\vskip .5truecm
\noindent
{\it Generalities.}

It is  well known that at the $SU(2)$-radius the BRST cohomology classes
for spin zero and ghost number zero are characterized by the operators ${\cal
O}_{u,n}$, with ${\cal O}_{0,0} \equiv 1 $, $x \equiv {\cal O}_{{1\over
2},{1\over 2}}$ and $y \equiv {\cal O}_{{1\over 2},-{1\over 2}}$ [4,5].
These operators
generate the chiral ground ring $C({\cal A}_L)$. The operators ${\cal
O}_{u,n}
= x^{u+n}\cdot y^{u-n}$ are precisely the polynomial functions on the $x-y$
plane with area form $\omega = dx\wedge dy$. Thus, the pair $({\cal A}_L,
\omega)$ is a two-dimensional symplectic manifold with the symplectic
two-form
$\omega$. A similar procedure done for the right side of ${\cal A}_L$ leads to
$({\cal
A}_R,\tilde \omega)$, where $\tilde \omega = d \bar x \wedge d \bar y$.

The spin zero and ghost number one states are

$$Y_{s,n}^+ = c V_{s,n}\cdot exp [\sqrt 2 \phi \mp s\phi \sqrt 2], \eqno(2.1)$$
with $s \geq 0$. These states are precisely the polinomial vector fields on
the $x-y$ plane that generate the area-preserving
diffeomorphisms. In terms of the symplectic form $\omega$ we have

$$ Y_{u+1,n}^{+i} = \omega^{ij} \partial_j {\cal O}_{u,n} = c W^{+i}_{u,n}.
\eqno(2.2)$$

The complementary vector fields are $a{\cal O}_{u,n}$ with $a{\cal O}_{u,n}(0)
= {1\over 2\pi i} \oint {dz\over z} a(z) {\cal O}_{u,n}(0)$, where $a$ is
defined in [5]. Then, the ghost number two states $aY_{s,n}^+$ $(s=u+1)$ are
the polynomial bivector fields ${\partial \over \partial x} \wedge {\partial
\over \partial y}$ on the $x-y$ plane.

If we denote the tangent bundle of the $x-y$ plane by $T{\cal A}$, with ${\cal
A}={\cal A}_L={\cal A}_R$, then the corresponding discrete states are
precisely the sections of the $\mu$-th order exterior algebra

$$ \pi^{\mu} : \ \Lambda^{\mu}T{\cal A} \to {\cal A} \eqno(2.3)$$
for $\mu=0,1,2$; we have $\Lambda^0T{\cal A} ={\cal C}$, $\Lambda^1T{\cal A}=
T{\cal A}$, etc. Here $\mu=0$ corresponds to the polinomial functions ${\cal
O}_{u,n}$; $\mu=1$ corresponds either to $Y^+_{s,n}$ or to $a{\cal
O}_{u,n}$, and $\mu=2$ does to
$aY^+_{s,n}$. Of course, there exists the dual version of this construction
where the exterior algebra is precisely the Grassmann-Cartan algebra
$\Omega(T^*{\cal A})$ (for details see Ref. [5].)

The exterior derivative corresponds to the $b_0$ operator which commutes  with
$\hat {\cal O} = {1\over 2\pi i} \oint {dw\over w} {\cal O}$; {\it i.e.} the
operator $[b_0,{\cal O}] = 0$ is BRST invariant. Only the zero forms
(or bivectors in the dual version) contribute to the central extension
$[{\partial \over \partial x},{\partial \over \partial y}] = \iota$ of the
polynomial area-preserving diffeomorphisms of the plane $SDiff({\cal A})$.

If $\alpha \in \Omega(T^*{\cal A})$, then

$$ b_0\alpha = d\alpha + \delta_{j=o} \alpha(0)\iota. \eqno(2.4) $$

As we deal with a flat and contractible ring (which is very much like as a
topological space) ring, a
closed one-form $\lambda$ on the plane $x-y$ is necessarilly exact {\it
i.e.},
$\lambda_{u,n} = d {\cal O}_{u,n}$ for some ${\cal O}_{u,,n}$. This
one-form $\lambda$ corresponds to the vector field

$$ Y^{+i}_{u+1,n} {\partial \over \partial x^i} = - (\partial_i {\cal O}_{u,n})
\omega^{ij}{\partial \over \partial x^j}, \eqno(2.5)$$
which is an area-preserving vector field derived from the hamiltonian function
${\cal O}_{u,n}$.

\vfill
\break
\noindent
{\it Ground Ring Manifold.}

 Given ${\cal A}_L$ and ${\cal A}_R$ we define  the {\it full quantum ground
ring manifold} ${\cal W}$ to be the product

$$ {\cal W} = {\cal A}_L \times {\cal A}_R. \eqno(2.6)$$

Which can be shown graphically

$$\matrix{&&{\cal A}_L \times {\cal A}_R \cr
&\rho \swarrow && \searrow \tilde \rho  \cr
{\cal A}_L &&&& {\cal A}_R \cr} \eqno(2.7)$$

\vskip 1truecm

\noindent
where $\rho$ and $\tilde \rho$ are the respective projections.

  To construct quantum field operators we have to combine left and
right movers. Thus, for the spin $(1,0)$ and spin $(0,1)$ currents we get

$$ {\cal J}_{s,n,n'} = W^+_{s,n}\cdot \bar {\cal O}_{s-1,n'} $$

$$ \bar{\cal J}_{s,n,n'} = {\cal O}_{s-1,n} \cdot \bar W^+_{s,n'}.
\eqno(2.8)$$

These operators act on the chiral ground ring operators according to the
rule

$${\cal J}({\cal O}(P)) = {1\over 2\pi i} \oint_C {\cal J}(z)\cdot {\cal
O}(P)$$

$$\bar {\cal J}(\bar{\cal O}(P)) = {1\over 2\pi i} \oint_C \bar{\cal
J}(z)\cdot \bar{\cal O}(P)\eqno(2.9)$$
and are precisely derivations of the full quantum ground ring.

It is an easy matter to see that due to the fact that both $({\cal
A}_L,\omega)$ and $(\tilde {\cal A}_R,\tilde \omega)$ are symplectic
manifolds and that ${\cal W} = {\cal A}_L \times {\cal A}_R$, the pair
$({\cal W}, \rho^*\omega - \tilde \rho^* \tilde \omega)$ is also a symplectic
manifold.

\vfill
\break

\centerline{\bf 3. Self-dual  Gravity Structures in 2d String Theory}

Consider the effective topological action for all discrete states of the $c=1$
string at the $SU(2)$ point. This action reads [5]:

$$ \int_{{\cal C}^2} \nu F\wedge F \eqno(3.1)$$
where $\nu$ is a scalar and $F$ is defined by the $U(1)$-gauge field $A$ on
${\cal C}^2$ as follows: $F=dA$. The $U(1)$ principal bundle $ p: {\cal W} \to
{\cal Q} = {\cal W}/{\cal H}$ produces a dimensional reduction of the action
(3.1)

$$S^*(\int_{{\cal C}^2/ \Gamma} \nu F\wedge F) = \int_{{\cal
C}^2/(\Gamma\otimes U(1))} \nu \cdot du\wedge da \eqno(3.2)$$
where $S$ is a section of the above bundle, ${\cal C}^2/\Gamma$ represents the
Kleinian singularities for the relevant subgroup $\Gamma$, $u$ is a function
on the 3-dimensional cone ${\cal Q}$ and $a$ is the pullback of an
abelian gauge field from ${\cal Q}$.
In [15] it is shown that the action (3.1) leads to the equation of motion,
which appears to be precisely the first heavenly equation

$$  \Omega_0\wedge \Omega_0 +  2 \omega \wedge \tilde \omega = 0,
\eqno(3.3a)$$
with $ \Omega_0 = \partial \bar \partial \Omega$, or equivalently

$$
\Omega_{x \bar x} \Omega_{y \bar y} - \Omega_{x \bar y} \Omega_{y \bar x} = 1.
\eqno (3.3b) $$

This shows an unexpected presence of self-dual gravity structures in the
string context.

One finds easily [15] that the element  of the full quantum ground ring

$$ \Omega = x \bar x + y \bar y \eqno (3.4)$$
satisfies the first heavenly equation (3.3b). Moreover if one
interprets $\Omega$ to be the K\"ahler function, then the corresponding
metric is flat.

On the other hand we can see that the element of the full quantum ground ring
$\Omega_{u,n,n'}(x,y,\bar x, \bar y) = {\cal O}_{u,n}\cdot \bar{\cal
O}_{u,n'}$ does not satisfy the first heavenly equation as it is, but a
"modified" one given by Q-Han Park in [16]

$$ \gamma(\Omega_{x\bar x} \Omega_{y\bar y} - \Omega_{x \bar y}\Omega_{\bar x
y}) + (1-\gamma) \Omega_{xy} = W(x,y) \eqno(3.5)$$
with $\gamma = 1$ and $W=0$. This case corresponds precisely to the
Topological Model (W-Z term only). That is

$$\Omega_{x \bar x} \Omega_{y \bar y} - \Omega_{x \bar y}\Omega_{y \bar x } =
0. \eqno(3.6)$$

\vskip 2truecm

\centerline{\bf 4. Self-dual Gravity Structures on Ground Ring
Manifolds}
 In the previous section  we have obtained the first heavenly equation of
the self- dual gravity from the action [15]

$$\int_{{\cal C}^4} \nu \cdot F\wedge F.$$

In this section, we  re-derive the same structure in a different way
making emphasis only on the fact that the full quantum ground ring
manifold ${\cal
W}$ is merely the product manifold ${\cal A}_L \times {\cal
A}_R$, and
that the chiral ground ring manifolds are symplectic manifolds with symplectic
two-forms $\omega = dx \wedge dy $ and $\tilde \omega = d\bar x \wedge d\bar
y$ respectively. For this end we use the construction of Ref.[1,12].

Consider the flat chiral complexified ground ring manifolds ${\cal A}_L^{{\cal
C}}$ and   ${\cal A}_R^{{\cal C}}$ and the complexified full
ground ring manifold $ {\cal W}^{{\cal C}} = {\cal A}_L^{{\cal C}}\times
{\cal A}_R^{{\cal C}}$. Since $({\cal A}_L^{{\cal C}}, \omega)$ and
$({\cal
A}_R^{{\cal C}}, \tilde \omega)$ are symplectic manifolds it is very easy to
show that  ${\cal W}^{{\cal C}}$ is also a symplectic manifold with symplectic
form $\rho^* \omega - \tilde \rho^* \tilde \omega$ ($\rho\ : {\cal
W}^{{\cal C}} \to {\cal A}_L^{{\cal C}} $ and $\tilde \rho \ : {\cal W}^{{\cal
C}} \to {\cal A}_R^{{\cal C}} $ are the respective projections).

Now, let  $T^r{\cal A}^{{\cal C}}_L$, $T^r{\cal A}^{{\cal C}}_R$ and $T^r{\cal
W}^{{\cal C}}$ be the $r$-th order holomorphic tangent bundles  of ${\cal
A}^{{\cal C}}_L$, ${\cal A}^{{\cal C}}_R$ and ${\cal W}^{{\cal C}}$,
respectively. Then we have the following sequence of projections for
${\cal A}^{\cal C}_L$:

$$ \rightarrow T^r{\cal A}^{\cal C}_L \to T^{r-1}{\cal A}^{\cal C}_L \to
...\to T^1{\cal A}^{\cal C}_L \to T^0{\cal A}^{\cal C}_L \cong {\cal A}^{\cal
C}_L . \eqno(4.1)$$
and similarly for ${\cal A}^{\cal C}_R$ and ${\cal W}^{\cal C}$.

Following [1,12] one can define functions ${\cal O}^{(\lambda)}_{u,n}$, $\bar
{\cal O}^{(\lambda ')}_{u,n'}$ and ${\cal V}_{u,n,n'}^{(\lambda)}$;  vector
fields $Y^{+(\lambda)}_{s,n}$, $\bar Y^{+(\lambda ')}_{s,n}$ as well as
${\cal
J}_{u,n,n'}^{(\lambda)}$,$\bar{\cal J}_{u,n,n'}^{(\lambda ')}$, and
differential forms $\omega ^{(\lambda)}$,$\tilde\omega ^{(\lambda)}$, (with
$\lambda = 0,1,2,...,r$) on the $r$-order holomorphic tangent (or cotangent)
bundles $T^r{\cal A}^{\cal C}_L$, $T^r{\cal A}^{\cal C}_R$ and $T^r{\cal
W}^{\cal C}$. For example

$$ {\cal O}^{(\lambda)}_{u,n}(j_r\circ \psi(0)) = {1\over \lambda
!}{d^{\lambda}({\cal O}_{u,n}\circ \psi) \over dt^{\lambda}} \mid _{t=0}$$

$$ {\cal V}^{(\lambda)}_{u,n,n'}(j_r\circ \psi(0)) = {1\over \lambda
!}{d^{\lambda}({\cal V}_{u,n,n'}\circ \psi) \over dt^{\lambda}}\mid_{t=0}
\eqno(4.2)$$
where $j_r(\psi) $ is the $r$-jet of the holomorphic curve $\psi$. Then,

$${\cal O}^{(\lambda)}_{u,n}(j_r\circ \psi (0))\cdot {\cal
O}^{(\lambda')}_{u,n'}(j_r\circ \psi (z)) = {1\over \lambda !}
{\partial^{\lambda}({\cal O}_{u,n}\circ \psi) \over \partial s^{\lambda}} \mid
_{s=0}\cdot {1\over \lambda ' !}{\partial^{\lambda'}({\cal O}_{u,n'}\circ
\psi) \over \partial t^{\lambda'}} \mid _{t=z}.\eqno (4.3)$$

 From Theorem 2 of [1] one can see that  $(T^r{\cal A}^{\cal C}_L,
\omega^{(\lambda)})$ and  $(T^r{\cal A}^{\cal C}_R, \tilde\omega^{(\lambda)})$
are symplectic manifolds. Therefore, we can define another symplectic
manifold
$T^r{\cal W}^{\cal C}$ with symplectic two form $\rho^*\omega^{(r)} - \tilde
\rho^*\tilde \omega ^{(r)}$. It is very easy to establish the  bundle
diffeomorphism

$$ q:T^r{\cal W}^{\cal C} \to \rho^*T^r{\cal A}^{\cal C}_L \oplus \tilde
\rho^* T^r{\cal A}^{\cal C}_R \eqno (4.4)$$

The following diagram summarizes our construction

$$\matrix {T^2{\cal W}^{\cal C}  &\to   &\rho^*T^2{\cal A}^{\cal
C}_L\oplus\tilde \rho^*T^2{\cal A}^{\cal C}_R &
\cr
\pi^2_1 \downarrow      &      & \downarrow \rho_1^2 & \searrow \tilde
\rho^2_1&
\cr
T{\cal W}^{\cal C} &&\rho^*T^2{\cal A}^{\cal C}_L   \to   &\tilde \rho
T^2{\cal A}^{\cal C}_R
\cr
\pi \downarrow && &&&\cr
{\cal W}^{\cal C} &&&& \cr} \eqno(4.5) $$

In order to show the existence of self-dual gravity structures on the
full quantum ground ring manifold we restrict ourselves to the case $r=2$.
As it is
mentioned in Ref.[12] $(\rho^* T^2 {\cal A}^{\cal C}_L, \rho^*\omega^{(2)} -
\tilde \rho^*\tilde \omega ^{(0)})$ and $(\rho^*T^2{\cal A}^{\cal C}_R,
\rho^*\omega^{(0)} - \tilde \rho^*\tilde \omega ^{(2)})$ are also symplectic
manifolds. Here

$$\omega^{(0)} = \omega = dx\wedge dy,$$
$$\tilde \omega^{(0)} = \tilde \omega = d\bar x\wedge d\bar y,$$

$$\omega^{(2)} = dx\wedge dy^{(2)} + dx^{(1)}\wedge dy^{(1)} + dx^{(2)}\wedge
dx,$$

$$\tilde \omega^{(2)} = d\bar x\wedge d\bar y^{(2)} + d\bar x^{(1)}\wedge
d\bar y^{(1)} + d\bar x^{(2)}\wedge d\bar x. \eqno (4.6)$$

  We need only to consider the manifold $\rho^*T^2{\cal A}^{\cal C}_L$.

To show the existence of a self-dual gravity structure in the 2d string
theory we use the arguments of  Refs. [1,12]

Consider the sequences

$$ {\cal N}\buildrel\rm i \over{\to} T^2 {\cal W}^{\cal
C}\buildrel\rm \pi^2_1 \over{\to} T{\cal W}^{\cal
C}\buildrel\rm \pi^1 \over{\to} {\cal W}^{\cal C}$$

$$ {\cal N}\buildrel\rm i \over{\to} T^2{\cal W}^{\cal
C}\buildrel\rm \rho^2_1 \over{\to} \rho^*T^2{\cal A}^{\cal
C}_L\buildrel\rm \rho^1 \over{\to} {\cal W}^{\cal C}\eqno(4.7)$$
where ${\cal N}$ is a horizontal lagrangian submanifold of both $T{\cal
W}^{\cal C}$ and $\rho^*T^2{\cal A}^{\cal C}_L$.

Let $\sigma : {\cal S} \to T^2{\cal W}^{\cal C},\ {\cal S} \subset {\cal
W}^{\cal C}$, be a holomorphic section such that $i({\cal N}) = \sigma_ ({\cal
S})$, where $i\ : \ {\cal N} \to T^2{\cal W}^{\cal C}$ is an injection.

 Let ${\cal L}({\cal S})$ be the set of all holomorphic sections of $T^2{\cal
W}^{\cal C} \to {\cal W}^{\cal C}$ such that $\pi_1^2\circ i({\cal N})$ and
$\rho^2\circ i({\cal N})$ are horizontal lagrangian submanifolds of $T{\cal
W}^{\cal C}$ and $\rho^*T^2{\cal A}^{\cal C}_L$, respectively. Thus ${\cal
L}({\cal S})$ is defined by the equations

$$\sigma_1^*(\omega^{(1)} - \tilde \omega^{(1)}) = 0, \ \ \ on\ \ {\cal
S}\subset {\cal W}^{\cal C}$$

$$\sigma_2^*(\omega^{(2)} - \tilde \omega^{(0)}) = 0, \ \ \ on\ \ {\cal
S}\subset {\cal W}^{\cal C}, \eqno(4.8)$$

\noindent
[Notice that we have omitted the pull-back $\rho^*$ and $\tilde \rho^*$
in these formulas.]

\noindent
where $\omega^{(1)} = dx \wedge dy^{(1)} + dx^{(1)}\wedge dy$,  $\tilde
\omega^{(1)} = d\bar x \wedge d\bar y^{(1)} + d\bar x^{(1)}\wedge d\bar y$,
$\sigma_1 = \pi_1^2 \circ \sigma$ and $\sigma_2 = \rho^2\circ \sigma$.

Now the problem arises of {\it how $\sigma \in {\cal L}({\cal S})$
determines  a
self-dual gravity structure on the open sets ${\cal S}$ of the full quantum
ground ring manifold ${\cal W}^{\cal C}$. }

The solution of this problem is given by the following

\noindent
{\it Theorem [1,12]:} Let  $\sigma \ : \ {\cal S} \to T^2{\cal W}^{\cal C}, \
{\cal S} \subset {\cal W}^{\cal C}$, be a holomorphic section. The triplet
$(\omega, \tilde \omega, \Omega_0) =(\sigma^* \omega^{(0)}, \sigma^*\tilde
\omega^{(0)}, \sigma^* \omega^{(1)})$ defines a self-dual structure
on ${\cal S}$ if and only if there exist a choice of a holomorphic section
$\sigma$ such that $\sigma^*(\omega^{(1)} - \tilde \omega^{(1)}) =0$ and
$\sigma^*(\omega^{(2)} - \tilde \omega^{(0)}) = 0$.  $\triangle$

\noindent
(For the proof see [1,12]).
 Thus, the desired self-dual gravity structures arise in a natural
manner from the mathematical structure of the quantum states in 2d string
theory.

Taking $\Omega_0 = \sigma^*_1 [dx\wedge dy^{(1)} + dx^{(1)} \wedge dy]$ we
arrive at the first heavenly equation

$$  \Omega_0\wedge\Omega_0 +  2\omega\wedge \tilde \omega = 0. $$
This can be extended to the cases with $r\geq 3 $. Then, by using the
projective
limit one can formulate the problem in terms of the infinite-dimensional
tangent
bundle $T^{\infty} {\cal W}^{\cal C} = \rho^*T^{\infty}{\cal A}^{\cal C}_L
\oplus \tilde \rho^* T^{\infty}{\cal A}^{\cal C}_R$ (for details see [1,12]).
It can be proved that given  $({\cal W}^{\cal C}, \rho^*\omega - \tilde
\rho^* \tilde \omega)$ a symplectic manifold $(T^{\infty}{\cal
A}^{\cal C}_L,
\omega_2(t))$ turns out to be a formal symplectic manifold, where $\omega_2
(t) =
\sum_{k=0}^{\infty} \pi_k^*\omega^{(k)}t^k$;  $t\in {\cal C}$, and $ \pi_k \
: \ T^{\infty}{\cal A}^{\cal C}_L \to T^k{\cal A}^{\cal C}_L$ is the natural
projection.

By the Proposition 2 of [1]  we observe that  $(T^{\infty} {\cal W}^{\cal C},
\omega(t))$ is a formal symplectic manifold with

$$ \omega(t)= t^{-1} \rho^* \omega_2(t) - t \tilde \omega_2(t^{-
1})\eqno(4.9)$$
where $t\in {\cal C}^*\equiv {\cal C} - \{0\}$ and $T^{\infty}{\cal W}^{\cal
C} = T^{\infty}{\cal A}^{\cal C}_L \times T^{\infty} {\cal A}^{\cal C}_R =
T^{\infty}({\cal A}^{\cal C}_L \times {\cal A}^{\cal C}_R) = \rho^*T^{\infty}
{\cal A}^{\cal C}_L \times \tilde \rho^* T^{\infty}{\cal A}^{\cal C}_R.$

\vskip 2truecm
\noindent
{\it Curved Twistor Construction on Full Quantum Ground Ring Manifolds}.

Consider the formal symplectic manifold $(T^{\infty}{\cal W}^{\cal C},
\omega(t))$. Since  ${\cal A}^{\cal C}_L$ and ${\cal A}^{\cal C}_R$ are
diffeomorphic, we have $T^{\infty}{\cal W}^{\cal C}= T^{\infty}{\cal
A}^{\cal
C}_L\times T^{\infty}{\cal A}^{\cal C}_R$. Define the holomorphic maps

$$ \hat{\cal D}= (D,I) : T^{\infty}{\cal A}^{\cal C}_L \times {\cal C}^* \to
T^{\infty}{\cal A}^{\cal C}_L \times {\cal C}^*, $$
where $I(t) =t^{-1}$ and the graph of the diffeomorphism $D$, $grD$, can be
identified with some local section $grD=\sigma ' :{\cal S} \to T^{\infty}{\cal
W}^{\cal C}$ such that $\sigma '^* \omega(t) = 0$. From Eq.(5.9), this last
relation holds if and only if

$$ D^*\omega_2(t^{-1}) = t^{-2} \omega_2(t).\eqno(4.10)$$

 Consider now a local section  $\sigma ''$ of the formal tangent bundle
$T^{\infty} {\cal W}^{\cal C} \to {\cal W}^{\cal C}$ on a open set ${\cal
S}\subset {\cal W}^{\cal C}$ such that $\sigma ''^* \omega(t) =0$. For $t\in
{\cal C}^*$

$$\sigma ''= (\Psi^A(t),\tilde \Psi^B(t^{-1})). \eqno(4.11)$$
Assume that $\Psi^A(t)$ and $\tilde \Psi^B(t^{-1})$ converge in some open
discs ${\cal U}_0$ and ${\cal U}_{\infty}$ ($0\in {\cal U}_0$ and $\infty \in
{\cal U}_{\infty})$ respectively, such that ${\cal U}_0 \cap {\cal
U}_{\infty} \not= \phi$. Consequently, the functions $\Psi^A: t\mapsto
\Psi^A(t)$ and $\tilde \Psi^B: s\mapsto \tilde \Psi^B(s)$ define local
holomorphic sections of the twistor space ${\cal T}$. Due to the condition
(4.10) defining the self-dual structure on the quantum ground ring manifold
${\cal W}^{\cal C}$ we get the transition functions for a global holomorphic
section $\Psi \in \tilde \Gamma({\cal T}).$ Thus one can  recuperate the
Penrose
twistor construction [17]. Of course the inverse process is also possible (see
[1,12]).

\vskip 2truecm

\centerline{\bf 5. Final Remarks }

In this work we have looked for self-dual gravity structures in 2d string
theory. This was motivated mainly by the works of Witten and
Zwiebach [5]
as well as Ghoshal {\it et al.} [15]. These papers provide a
"physical" approach
that shows an unexpected presence of self-dual gravity structures in
string
theory. Goshal {\it et al.} [15] have found these structures by looking
for the
solutions to the equations of motion derived from the action (3.1). One class
of solutions in particular implies the existence of self-dual gravity
structures. Here we
have proposed the purely geometric ("mathematical") approach based on the
symplectic geometry of the chiral ground ring manifold. As we have shown,
this
geometry leads directly to self-dual structures in 2d string theory. Now, the
natural question arises about what is the relation  between the "physical"
and the
"mathematical" approaches. If Ghoshal {\it et al.} conjecture [15], which
states  that the dynamics of the states in 2d string theory  is given by the
self-dual gravity structure, is true, our geometric  approach might be more
convenient.

\vskip 2truecm
\centerline{\bf Acknowlegements}
One of us (M.P.) is grateful to the staff of Departamento de F\'{\i}sica at
CINVESTAV for warm hospitality. We are grateful to Dr. R. Capovilla, F. Larios
and the referee for useful suggestions. H. G-C. whish to thank CONACyT
and SNI for support.

\vskip 3truecm
\centerline{\bf Appendix. The Chiral Ground Ring Structure.}

In this appendix, we review the basic arguments which lead to the {\it Chiral
Ground Ring Structure}, as proposed by [4]. This structure arises in 2d string
theory. The bosonic
string theory with 2d target spacetime involves Liouville gravity coupled to
some conformal field theory (CFT) with central charge $c=1$. In the absence of
a cosmological constant, Liouville gravity and the CFT decouple. This gives
the world-sheet Lagrangian

$$ {\cal L} = {1\over 8 \pi} \int_{\Sigma} d^2 x \sqrt h (h^{ij} \partial _i X
\partial_j X + h^{ij} \partial_i \phi \partial_j \phi) -{1\over 2 \pi \sqrt 2}
\int_{\Sigma} d^2 x \sqrt h \cdot \phi R^{(2)}, \eqno (A1) $$
where $\Sigma$ is a Riemann surface, $h$ the world-sheet metric, $X$  a
bosonic field, $\phi$ a Liouville field and $R^{(2)}$ is the Ricci scalar.

The Lagrangian (A1) contains an infinite number of discrete states in addition
to the tachyonic state $V_p = exp (ipX)$ with $X$ as a free field. The presence
of these additional discrete states was discovered for the first time in
the study  of $c=1$ matrix models.

Introducing a $SU(2)$ symmetry on the states of (A1), these discrete
states arise in a natural way. Concretely, using the theory
compactified at
the $SU(2)$ radius, the momenta $p$ of $X$ take the discrete values
$p=n/\sqrt 2$, $n\in {\cal Z}$.

One can see that the conformal operator $exp(isX/\sqrt 2)$
belongs to a multiplet of a $SU(2)$ representation corresponding to the
highest weight. The other members of this multiplets are the operators
$V_{s,n}$ such that $V_{s,s} = exp(isX/\sqrt 2)$ and $V_{s,-s} = exp(-isX/\sqrt
2)$. The operator with $s= |n|$ corresponds precisely to the tachyon
operator. (some states remain for $|n|< s$)

Now, introducing the ghost fields $b$ and $c$ of spin  2 and -1 respectively,
one can  construct the spin 0 BRST invariant primary fields of ghost number 1

$$ Y^{\pm}_{s,n} = c W^{\pm}_{s,n} \eqno(A2)$$
from the spin 1 fields

$$ W^{\pm}_{s,n} = V_{s,n} exp(\sqrt 2 \phi \mp s \phi \sqrt 2). \eqno (A3)$$
The operators $W^{\pm}_{s,n}$ have momentum $(n,i(-1\pm s))\cdot 2$.

The composite  states $Y^{\pm}_{s,n}$ with $|n|< s$ have partners at an
adjoining value of the ghost number 0 or 2. Here we would like to consider
only those of ghost number 0. These will be partners of the positive part of
(A2) $Y^+_{s,n}$. Redefining $s= u+1$, these partner states are ${\cal
O}_{u,n}$ and have momentum $(n,iu)\cdot \sqrt 2$. The $u$'s take the values
$0,{1\over 2}, 1,$ etc, and $n=u,u-1,..., -u$.

The correspondence  between states with ghost number 1 and spin 1 and those
with ghost number 0 and spin 0 is

$$ Y^+_{1,0} = c \partial X \Leftrightarrow  {\cal O}_{0,0} = 1 $$

$$ Y^+_{{3\over 2},{1\over 2}} \Leftrightarrow x \equiv {\cal O}_{{1\over 2},
{1\over 2}} = (cb + {i\over 2}(\partial X - i
\partial \phi)) exp [i(X + i \phi)/2] $$

$$  Y^+_{{3\over 2},-{1\over 2}} \Leftrightarrow y \equiv {\cal O}_{{1\over
2}, -{1\over 2}} = (cb - {i\over 2}(\partial X + i \partial \phi)) exp - [i(X
- i \phi)/2]. \eqno(A4)$$
These states can be constructed  using the BRST analysis.

Combining the operators (A3) and (A4) one can define the {\it quantum field
operators} of spin (1,0) and (0,1) to be, respectively,

$$ {\cal J}_{s,n,n'} = W^+_{s,n}\cdot \bar {\cal O}_{s-1,n'} $$

$$ \bar{\cal J}_{s,n,n'} = {\cal O}_{s-1,n} \cdot \bar W^+_{s,n'}.
\eqno(A5)$$
where the bar represent complex conjugation.

These operators generate  a Lie algebra of symmetries, namely the Lie algebra
of {\it vo\-lume pre\-serving diffeomorphisms} of a 3 dimensional
algebraic variety defined by

$$ a_1a_2 - a_3a_4 = 0 \eqno(A6)$$
where $a_1 = x\bar x$, $a_2 = y\bar y$, $a_3= x \bar y$ and $a_4 = \bar x y$.

The {\it chiral ground ring} ${\cal A}_L$ structure come from the pair
$\{x,y\}$ ( similarly for the right part taking $\{\bar x, \bar y\}$).
${\cal
A}_L$ defines a ring structure under the usual operator product
expansion
[4]. The chiral symmetries  of the Lagrangian (A1) form the group of
diffeomorphisms preserving the area of the plane generated by $\{x,y\}$. This
group is denoted by $SDiff({\cal A}_L)$.

In this way, the existence of both the set of states (A4) and the
chiral ground ring
structure implies the existence of a $W_{\infty}$-symmetry in 2d string
theory.

Further, the discussion in Ref.[5] shows that the pair $({\cal A}_L,\omega)$
is a symplectic manifold, with symplectic 2-form $\omega = dx
\wedge dy$.

\vfill
\break

\centerline{\bf References}

\item{1.} C.P. Boyer and J.F. Pleba\'nski, {\it J. Math. Phys.} {\bf 26},
(1985) 229.

\item{2.} E. Witten, {\it Phys. Rev.}\ {\bf D46}\ (1992) 5467.

\item{3.} B. Zwiebach, {\it Nucl. Phys.} {\bf B390} (1993) 33.

\item{4.} E. Witten, {\it Nucl. Phys.}\ \ {\bf B373}\ (1992) 187.

\item{5.} E. Witten and B. Zwiebach, {\it Nucl. Phys.} {\bf B377} (1992) 55.

\item{6.} S. Kachru, {\it Mod. Phys. Lett. A} {\bf 7} (1992) 1419.

\item{7.} J.L. Barb\'on, {\it Int. J. Mod. Phys.} {\bf A7} (1992) 7579.

\item{8.} K. Takasaki, "Area-preserving diffeomorphisms and nonlinear in\-te\-
grable sys\-tem" Pre\-print KUCP-0039, October 1991; {\it Phys. Lett.} {\bf
B285} (1992) 187.

\item{9.} J.F. Pleba\'nski, {\it J. Math. Phys.} {\bf 16} (1975) 2395.

\item{10.} B. H. Lian and G.J. Zuckerman, {\it Commun. Math. Phys.} {\bf 154},
(1993) 613.

\item{11.} C.P. Boyer and J.F. Pleba\'nski, {\it J. Math. Phys.} {\bf 18},
(1977) 1022.

\item{12.} C.P. Boyer, in {\it Nonlinear Phenomena}, K.B. Wolf (ed) Lecture
Notes in Physics vol. 189 Springer-Verlag (1983).

\item{13.} J.F. Pleba\'nski and I. Robinson, {\it Phys. Rev. Lett.} {\bf 37}
(1976) 493.

\item{14.} C.P. Boyer, J.D. Finley III and J. F. Pleba\'nski, "Complex
general relativity, ${\cal H}$ and ${\cal HH}$ spaces- a survey" Instituto de
Investigaciones en Matem\'aticas Aplicadas y en Sistemas. UNAM (1978).

\item{15.} D. Ghoshal, D.P. Jatkar and S. Mukhi, {\it Nucl.Phys.} {\bf B395}
(1993) 144.

\item{16.} Q-Han Park, {\it Int. J. Mod. Phys.} {\bf A7} (1992) 1415.

\item{17.} R. Penrose, {\it Gen. Rel. Grav.} {\bf 7} (1976) 31.

\bye